\pdfoutput=1
\documentclass[%
 reprint,
superscriptaddress,
nofootinbib,
 amsmath,amssymb,
 aps,
pra,
]{revtex4-1}
\usepackage{graphicx}
\usepackage{bm}
\usepackage{hyperref}
\usepackage[mathlines]{lineno}
\usepackage[utf8]{inputenc}
\usepackage{amsmath,amssymb}
\usepackage[T1]{fontenc} 
\usepackage{microtype} 
\usepackage[english]{babel} 

\usepackage{booktabs} 

\usepackage{graphicx}
\usepackage{amssymb}
\usepackage{braket,mleftright}
\usepackage{empheq}
\usepackage{subfigure}
\usepackage{stackrel}
\usepackage{soul}
\usepackage{blkarray}
\usepackage{multirow}
\usepackage{amsmath}
\usepackage{physics}
\usepackage{amsfonts}
\usepackage{bm}
\usepackage{bbold} 
\usepackage{color}
\usepackage{natbib}
\setcitestyle{square, comma, numbers,sort&compress, super}

\newcommand{\beq}{\begin{equation}}
\newcommand{\eeq}{\end{equation}}
\newcommand{\bse}{\begin{subequations}}
\newcommand{\ese}{\end{subequations}}
\newcommand{\bea}{\begin{eqnarray}}
\newcommand{\eea}{\end{eqnarray}}

\usepackage[utf8]{inputenc} 
\usepackage{amsmath}

\usepackage{enumitem} 
\setlist[itemize]{noitemsep} 

\usepackage{hyperref}
\usepackage{cleveref}
\usepackage{braket}
\usepackage{verbatim} 
\begin{document}


\title{Information backflow as a resource for entanglement}

\author{Nicol\'{a}s Mirkin}
\email[Corresponding author:]{mirkin@df.uba.ar}
\affiliation{%
Departamento de F\'{i}sica “J. J. Giambiagi” and IFIBA, FCEyN, Universidad de Buenos Aires, 1428 Buenos Aires, Argentina
}%
\author{Pablo Poggi}
\affiliation{%
Departamento de F\'{i}sica “J. J. Giambiagi” and IFIBA, FCEyN, Universidad de Buenos Aires, 1428 Buenos Aires, Argentina
}%
\affiliation{Center for Quantum Information and Control, University of New Mexico, MSC07-4220, Albuquerque, New Mexico 87131-0001, USA}
\author{Diego Wisniacki}

\affiliation{%
Departamento de F\'{i}sica “J. J. Giambiagi” and IFIBA, FCEyN, Universidad de Buenos Aires, 1428 Buenos Aires, Argentina
}%

\date{March 18, 2019}%

\begin{abstract}
The issue of whether non-Markovianity (NM) could be considered as a resource in quantum information has been a subject of intense debate for the last years. Recently, a simple mechanism was proposed in which one of the main features of NM, the backflow of information from the environment to the system, represents a fundamental and quantifiable resource for generating entanglement within an open quantum system coupled to a finite and small environment \mbox{\href{https://doi.org/10.1103/PhysRevA.99.020301}{[N. Mirkin, P. Poggi and D. Wisniacki, Phys. Rev. A, 99(2), 020301(R)]}}. In this work, we extend the universality of this resource mechanism by studying a completely different and more general scheme where the system is coupled to an infinite structured reservoir. Under both setups, we show that the degree of NM univocally determines the optimal degree of entanglement reachable by controlling the open system. This result reveals the universality of a quantitative relation between entanglement and NM by using quantum optimal control. 
\end{abstract}

\maketitle

\section{\label{Section-Intro}Introduction}
Quantum technologies promise to lead a new revolution in the so-called Information Age. But for the prosperous development of these quantum technologies it is critical to control a big variety of quantum systems with high efficiency in the shortest time that is physically possible \cite{bib:intro1,bib:intro2,bib:intro3,bib:intro4}. One of the main difficulties under this context, is the capability of manipulating realistic quantum systems which are unavoidably subject to an interaction with the environment. Specifically, the question of how to deal with the detrimental effects of the environment such as decoherence is one of the most fundamental challenges in the area  \cite{bib:intro_deco,bib:petru}. Nevertheless, over the last years, the possibility of exploiting the environment as a resource for control has opened a new door in the manipulation of open quantum systems \cite{bib:env_resource0,bib:env_resource1,bib:env_resource2,bib:env_resource3}. In particular, the so-called non-Markovianity (NM), mainly associated with non-divisibility of quantum maps and with a backflow of information from the environment to the system \cite{bib:medida1,bib:medida2}, has been pointed as beneficial in a diverse set of settings, including the protection of entanglement properties \cite{bib:manis2,bib:lofranco, bib:manis3,bib:plenio,bib:lofranco3}, the decrease of quantum speed limit \cite{bib:intro_speedup,bib:intro_speedup2,bib:intro_speedup3,bib:intro_speedup4}, the implementation of quantum algorithms \cite{bib:algorithm} or even in the power of quantum thermal machines \cite{bib:termo1,bib:termo2}.

However, while the best definition for quantum NM is still matter of huge controversy in the literature \cite{bib:medida3,bib:nmreport1,bib:nmreport2,bib:pineda,bib:modi}, a fully accepted resource theory for such a quantity has not been formalized yet \cite{bib:nm_resource}. Under this resource framework, one may be tempted to think that a certain amount of NM in the system dynamics would allow to accomplish a particular task that would be impossible to achieve in the absence of it. With this idea in mind, in a recent work we have proposed to use optimal control techniques as a tool to search for a quantitative relation between the amount of success obtained for a particular set of entangling tasks and the degree of NM present in the dynamics. For this, we focused our study on a physical scenario composed of $N$ non-interacting subsystems coupled to the same non-Markovian environment, where the task to accomplish consisted on driving this from a separable initial state to an entangled target state  \cite{bib:mirkin_ent}. There, by studying the particular case of a spin star configuration, where $N$ non-interacting central spins are coupled to the same finite and small set of environmental spins, we have shown that the optimal degree of controlled entanglement reached by the optimization was a direct function of the original amount of NM of the system dynamics. 

Despite the strength and interesting nature of this last result, a fundamental question regarding what happens in more complicated and realistic environments such as infinite structured reservoirs remained unanswered. As a consequence, in this work we validate and extend the universality of the resource mechanism by studying a completely different scheme, where two non-interacting two-level atoms are coupled to the same infinite structured reservoir, i.e. a leaky cavity composed of $M$-harmonic oscillators. By first deriving the full time-evolved density matrix of the reduced system, we show in this new platform that the degree of NM univocally determines the degree of success of the entangling protocol considered, as observed in the case of the finite size environment studied previously. Furthermore, for both scenarios, we explore other control schemes in which we vary the degree of control available over each subsystem, and discuss the different roles played by NM in each particular case.

The fact that the same quantitative relation between NM and entanglement is fulfilled in two radically different systems, strongly suggests the existence of a universal mechanism for exploiting one of the most typical features of NM, i.e. the backflow of information from the environment back to the system, to generate entanglement. 
 
The manuscript is organized as follows. In Section II we present both systems under which we construct our results and we also discuss the measure of NM that is used. With respect to Section III, we address the interplay between NM and entanglement by using quantum optimal control techniques to drive the open quantum system from an initial separable state to a final target entangled state. Finally, we conclude in Section IV with a final discussion and our future perspectives. 

\section{\label{Section-Systems}Systems and Methods}
\subsection{Two atoms in a leaky cavity}
We consider two non-interacting two-level atoms, each one coupled to the same zero-temperature bosonic reservoir composed by a set of $M$-harmonic oscillators \cite{bib:manis2,bib:lofranco4}. The total microscopic Hamiltonian which describes the dynamics is of the form
\begin{equation}
\begin{split}
    H & =  \ H_{S} + H_{E} + H_{int} \\  & =\omega_{1}(t)\sigma_{+}^{(1)}\sigma_{-}^{(1)} + \omega_{2}(t)\sigma_{+}^{(2)}\sigma_{-}^{(2)} + \sum_{k=1}^{M}\omega_{k}b^{\dagger}_{k}b_{k} 
    \\ & + (\alpha_{1}\sigma_{+}^{(1)}+\alpha_{2}\sigma_{+}^{(2)})\otimes \sum_{k=1}^{M}g_{k}b_{k}+h.c.
\end{split}
\label{eq_hamilt}
\end{equation}
where $\sigma_{\pm}^{(i)}=\dfrac{1}{2}(\sigma_{x}^{(i)}\pm i \sigma_{y}^{(i)})$ and $\sigma_{j}^{(i)}$ ($j=x,y,z$) are the Pauli matrices of the atom $\textit{i}$ ($i=1,2$), $g_{k}$ is the coupling constant to the k-th mode of the bath, $b_{k}$ and $b_{k}^{\dagger}$ the usual annihilation and creation operators, $\alpha_{i}$ is a dimensionless constant that measures the interaction with the reservoir, and $\omega_{i}(t)$ is the time dependent energy difference between states $\ket{1}$ and $\ket{0}$ of the atom $\textit{i}$, which we will assume to be of the form
\begin{equation}
    \omega_{i}(t)=\omega_{0}+\epsilon_{i}(t).
\end{equation}
We consider $\epsilon_{i}(t)$ to be an arbitrary driving field over the atom $\textit{i}$, with which we intend to drive the open system from an initial separable state to an entangled target state. We now proceed to derive an equation for the reduced dynamics of both atoms. Starting with an initial state of the form
\begin{equation}
\ket{\phi(0)}=\left(C_{01}\ket{10}+C_{02}\ket{01} \right)\otimes_{k}\ket{0_{k}},
\end{equation}
the time evolution of the total system is given by
\begin{equation}
\begin{split}
\ket{\phi(t)} &=\ C_1(t)\ket{10}\ket{0_{B}}+C_2(t)\ket{01}\ket{0_{B}} \\ & +\sum_{k}C_{k}(t)\ket{00}\ket{1_{k}},
\end{split}
\end{equation} 
being $\ket{1_{k}}$ the state of the reservoir with only one excitation in the k-th mode. Following the procedure put forward in the Appendix, we can derive the following coupled differential equations for $C_1(t)$ and $C_2(t)$: 
\begin{equation}
    \Ddot{C}_{1}+(\lambda-i\epsilon_{1}(t))\dot{C}_{1}+\alpha_{1}\dfrac{\gamma_{0}\lambda}{2}\Big(\alpha_{1}C_{1}+\alpha_{2}e^{i(v_{1}-v_{2})}C_{2}\Big)=0
\label{eqc1}
\end{equation}
and for symmetry
\begin{equation}
        \Ddot{C}_{2}+(\lambda-i\epsilon_{2}(t))\dot{C}_{2}+\alpha_{2}\dfrac{\gamma_{0}\lambda}{2}\Big(\alpha_{2}C_{2}+\alpha_{1}e^{-i(v_{1}-v_{2})}C_{1}\Big)=0.
\label{eqc2}
\end{equation}
In the Eqs. above, $\gamma_{0}$ refers to the coupling between the system and the bath, $\lambda$ determines the width of the spectral density of the bath, $\epsilon_{i}(t)$ is the control field over the atom $\textit{i}$ ($i=1,2$) and $v_{i}(t)=\int_{0}^{t}ds \omega_{i}(s)$. Note that in the case in which we are driving with $\epsilon_{1}(t)=\epsilon_{2}(t) \, \forall t$, since $v_{1}(t)-v_{2}(t)=\int_{0}^{t}ds\Big(\epsilon_{1}(s)-\epsilon_{2}(s) \Big)$, then this last factor is not present in the dynamics. At the same time, is interesting to point out that the limit of $\lambda \rightarrow 0$ corresponds to the physical situation in which the spectral density is a delta function (i.e. the system being coupled to just one mode of the cavity). Is easy to see that in this case the differential equations that govern the dynamics of both atoms become decoupled and so the control field may not be able to generate any entanglement between them. The density matrix can be written as \cite{bib:manis2, bib:lofranco4}
\begin{equation}
    \rho(t)=\begin{pmatrix}
0 & 0 & 0 & 0 \\
0 & \abs{C_{1}(t)}^{2} & C_{1}(t)C^{*}_{2}(t) & 0 \\
0 & C^{*}_{1}(t)C_{2}(t) & \abs{C_{2}(t)}^{2} & 0 \\
0 & 0 & 0 & 1-\abs{C_{1}(t)}^{2}-\abs{C_{2}(t)}^{2}
\end{pmatrix},  
\end{equation}
where $C_1(t)$ and $C_2(t)$ are given by solving numerically Eqs. (\ref{eqc1}) and (\ref{eqc2}), respectively. Finally, as in any optimal control problem, we also need to define the functional which we intend to maximize. In our case, the functional chosen is the concurrence, which for the system of interest is given by \cite{bib:manis2}    
\begin{equation}
\mathcal{C}(t)=2|C_1(t)C_{2}^{*}(t)|.    
\label{eq_conc}
\end{equation}
Is simple to check that in the case in which $C_{1}(T)=C_{2}(T)=1/\sqrt{2}$, the entanglement quantified by the concurrence is maximal.

\subsection{Spin star configuration}
To strengthen the generality of the main result of our work and to extend the analysis made in \cite{bib:mirkin_ent}, let us also take as a physical model a spin star configuration, where two non-interacting central spin-$\frac{1}{2}$ particles are surrounded by a set of $N-2$ likewise environmental particles \cite{bib:spinstar1,bib:spinstar2,bib:spinstar3,bib:spinstar4}. The $l$-th central spin is coupled to the $k$-th environmental spin via the coupling constant $A_{k}^{(l)}$ (\textit{$l=1,2$}). However, to simplify the model, let us assume that the central spins are equally coupled to the environmental spins, i.e. $A_{k}^{(l)}=A_{k'}^{(l')}=A$ with $k \neq k'$ and $l \neq l'$. Moreover, let us consider that the central spins are being controlled via two time-dependent control fields in the $\hat{y}$ direction. The Hamiltonian that governs this model is given by
\begin{equation}
\begin{split}
    H & = H_{0}+H_{C}(t) \\
    & = \dfrac{\omega_{0}}{2}\sigma_{z}^{(1)}+\dfrac{\omega_{0}}{2}\sigma_{z}^{(2)}
    +\sum_{k=1}^{N-2} \left( A \, \vec{\sigma}^{(1)}.\vec{\sigma}^{(k+2)} \right) \\
    &    +\sum_{k=1}^{N-2} \left( A \, \vec{\sigma}^{(2)}.\vec{\sigma}^{(k+2)} \right)
    +\epsilon_{1}(t)\sigma_{y}^{(1)}+\epsilon_{2}(t)\sigma_{y}^{(2)},
\end{split}
\end{equation}
where $H_{0}$ plays the role of the free Hamiltonian and $H_{C}(t)$ is the control Hamiltonian. The operators $\vec{\sigma}^{(l)}$ and $\vec{\sigma}^{(k)}$ are the Pauli operators of the $l$-th central spin and the $k$-th environmental spin, respectively, and the quantity $\epsilon_{l}(t)$ is the control field over the $l$-th central spin. We pretend to optimize those fields such as to drive the open system from a separable initial state to an entangled target Bell-state of the form  $\ket{\Phi^{(+)}}=\dfrac{1}{\sqrt{2}}(\ket{00}+\ket{11})$. In this particular case, the functional we intend to maximize is the state fidelity defined as $\mathcal{F}_{state}=|\bra{\psi_{targ}}\ket{\psi(T)}|^{2}$ \cite{bib:qutip}. However, to be consistent, after optimizing the state fidelity we will compute the concurrence of this optimal state and relate this last quantity to the degree of NM of the system dynamics. The concurrence is defined as $\mathcal{C}(t)=max\{0,\sqrt{\lambda_{1}}-\sqrt{\lambda_{2}}-\sqrt{\lambda_{3}}-\sqrt{\lambda_{4}}\}$, where $\{\sqrt{\lambda_{i}}\}$ are the eigenvalues of the matrix $R=\rho(\sigma_{y}^{A}\otimes\sigma_{y}^{B})\rho^{*}(\sigma_{y}^{A}\otimes\sigma_{y}^{B})$, with $\rho^{*}$ denoting the complex conjugate of $\rho$ and $\sigma_{y}^{A/B}$ being the Pauli matrices of central spins A and B, respectively. 

\subsection{Non-Markovianity measure}
Considering that the difficulty in establishing a single measure for quantifying non-Markovian effects may be based on the fact that what is called NM is actually something encompassing different aspects of open quantum dynamics, in this work we follow a pragmatical approach and exclusively focus on the backflow of information feature, understanding the different measures as descriptions of different properties of open quantum systems \cite{bib:manis}. 
In this sense, one of the main approaches to quantify this effect was developed by Breuer, Laine and Piilo (BLP) \cite{bib:medida1}, who based their measure in the revivals of distinguishability between quantum states during the dynamics. The BLP criterion states that a quantum map is non-Markovian if there exists at least a pair of initial states $\rho_{1}(0)$ and $\rho_{2}(0)$ such that the distinguishability between them increases during some interval of time. The distinguishability can be quantified by the trace distance, which is defined as $D(\rho_{1},\rho_{2})=\dfrac{1}{2}||\rho_{1}-\rho_{2}||$ and where $||A||=tr(\sqrt{A^{\dagger}A})$. The fact that the states $\rho_{1}(t)$ and $\rho_{2}(t)$ are becoming momentarily more distinguishable implies that information has flowed from the environment back to the system. Therefore, if during some interval of time we have that
\begin{equation}
    \sigma(\rho_{1}(0),\rho_{2}(0),t)=\dfrac{d}{dt}D(\rho_{1}(t),\rho_{2}(t)) > 0,
\label{distinguibilidad}
\end{equation}
then we are in presence of non-Markovian dynamics. This idea can also be extended to define a measure of the degree of NM in a quantum process via 
\begin{equation}
    \mathcal{N}_{BLP}=\max\limits_{{\lbrace\rho_{1}(0),\rho_{2}(0)\rbrace}} \int_{0, \sigma >0}^{T}\sigma \left (\rho_{1}(0),\rho_{2}(0),t'\right) dt',
    \label{BLP}
\end{equation}
where T stands for the final evolution time of the process considered. In order to compute the BLP measure for the degree of NM, we take as initial orthogonal states $\rho_{1}(0)=\ket{10}\bra{10}$ and $\rho_{2}(0)=\ket{01}\bra{01}$ \cite{bib:manis,bib:wissmann}. Finally, let us also note that NM here is quantified for a restricted time interval, due to considering a finite evolution time for the control protocol, which may be varied.

\section{Non-Markovianity and entanglement}
\begin{figure}[!htb]
\begin{center}
\includegraphics[scale=0.46]{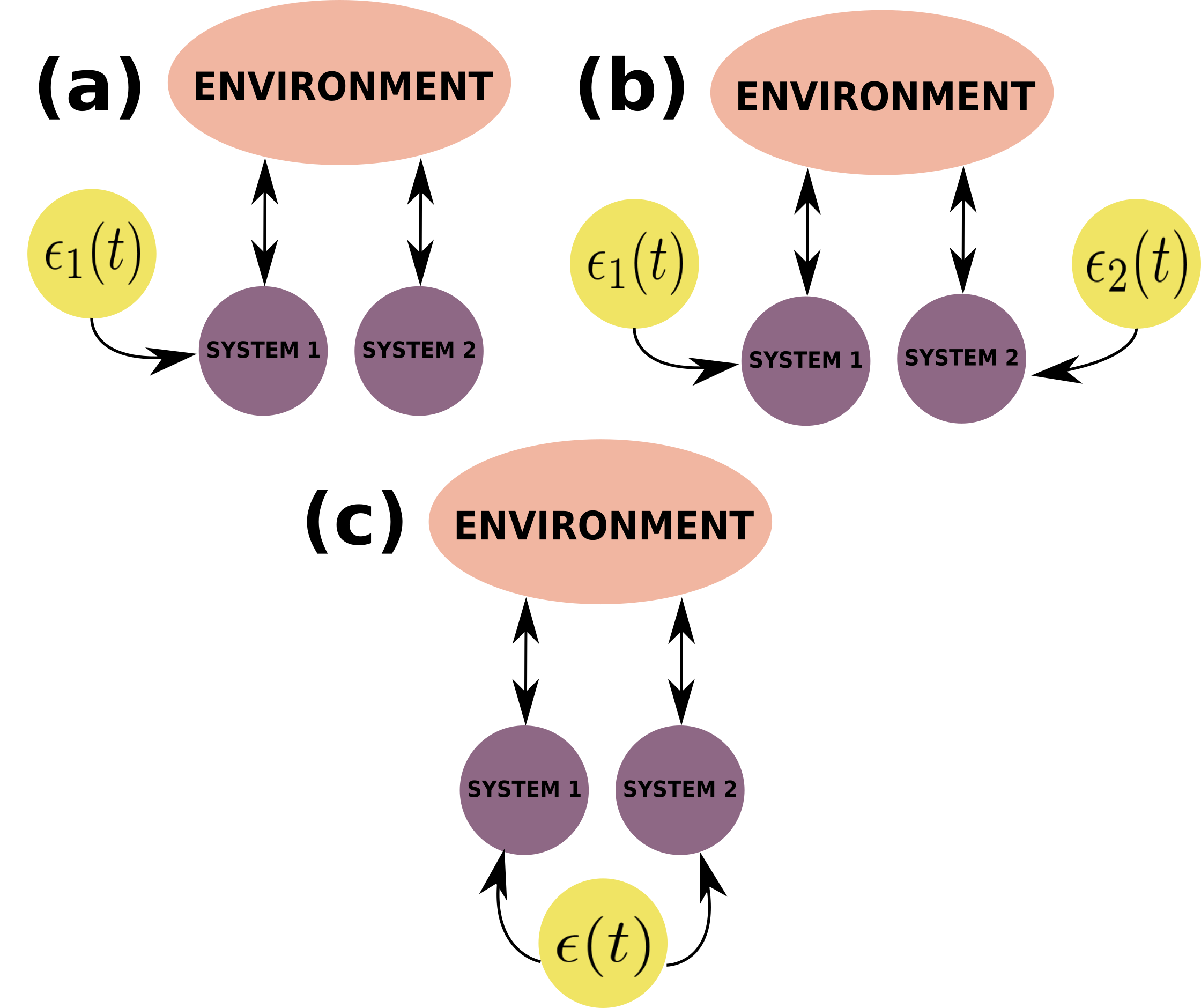}
\begin{footnotesize}
\caption{The three different entangling protocols considered under non-Markovian dynamics. (a) \textit{Single Addressing (SA)}: we can just access and control one of the systems. (b) \textit{Double Addressing (DA)}: we control both systems with different fields. (c) \textit{Global Addressing (GA)}: the same field is simultaneously controlling both of the systems.}
\label{nm_ent}
\end{footnotesize}
\end{center}
\end{figure}

In order to seek for a quantitative interplay between entanglement and NM, we study three different control protocols for entangling the non-interacting systems and relate them with the original amount of NM in the system dynamics. The first protocol is named \textit{Single Addressing} since just one of the subsystems can be accessed and controlled via $\epsilon_{1}(t)$. In the second protocol we control each subsystem with a different field $\epsilon_{1}(t)$ and $\epsilon_{2}(t)$ (\textit{Double Addressing}), while in the last protocol both subsystems are being controlled by the same field $\epsilon(t)$ (\textit{Global Addressing}). In Fig. \ref{nm_ent} we schematically show all these protocols. 
       
\renewcommand{\figurename}{Figure} 

Let us first analyze qualitatively all the entangling protocols under consideration. In first place, in the case of \textit{Single Addressing} control is interesting to note that the field $\epsilon_{1}(t)$ just delivers information to system 1 ($S_{1}$), but in order to maximize the concurrence needs to hand in some way that information to system 2 ($S_{2}$). However, as can be seen from the panel (a) in Fig. \ref{nm_ent}, such as to give that information to $S_{2}$ and so being able to control it, a flow of information from the environment ($E$) back to the system seems to be completely necessary. This may only occur under the presence of non-Markovian dynamics, since there a new channel of information, i.e. $\epsilon_{1}(t) \rightarrow S_{1} \rightarrow E \rightarrow S_{2}$, is enabled. On the other hand, in the case of \textit{Double Addressing}, the situation is quite similar but where two different fields, each one coupled to one of the subsystems, need to deliver information to the other subsystem. As before, this seems possible exclusively because of the features of NM and the existence of a backflow of information from the environment to the system. Under this frame, in addition to the flows $\epsilon_{1}(t) \rightarrow S_{1}$ and $\epsilon_{2}(t) \rightarrow S_{2}$ we are enabling two extra flows of information, i.e. $\epsilon_{1}(t) \rightarrow S_{1} \rightarrow E \rightarrow S_{2}$ and $\epsilon_{2}(t) \rightarrow S_{2} \rightarrow E \rightarrow S_{1}$. Considering that here we have more information channels to perform the control task in comparison to the \textit{Single Addressing} platform, we should expect a better degree of success under this \textit{Double Addressing} scheme. Finally, in the case of \textit{Global Addressing} control, there is one unique field that controls both of the systems and delivers the same information to them. Thus, there seems to be no gain in the presence of non-Markovian dynamics since the information that could flow from $\epsilon(t) \rightarrow S_{1} \rightarrow E \rightarrow S_{2}$ is the same that flows from $\epsilon(t) \rightarrow S_{2}$.

In the following subsections we test this qualitative argument by showing a quantitative analysis, both for the case of an infinite structured non-Markovian environment composed of harmonic oscillators and for the case of a finite small-size environment formed by a set of environmental spins.   

\subsection{Infinite structured environment: Two atoms in a leaky cavity}
\begin{figure}[!htb]
\begin{center}
\includegraphics[scale=0.72]{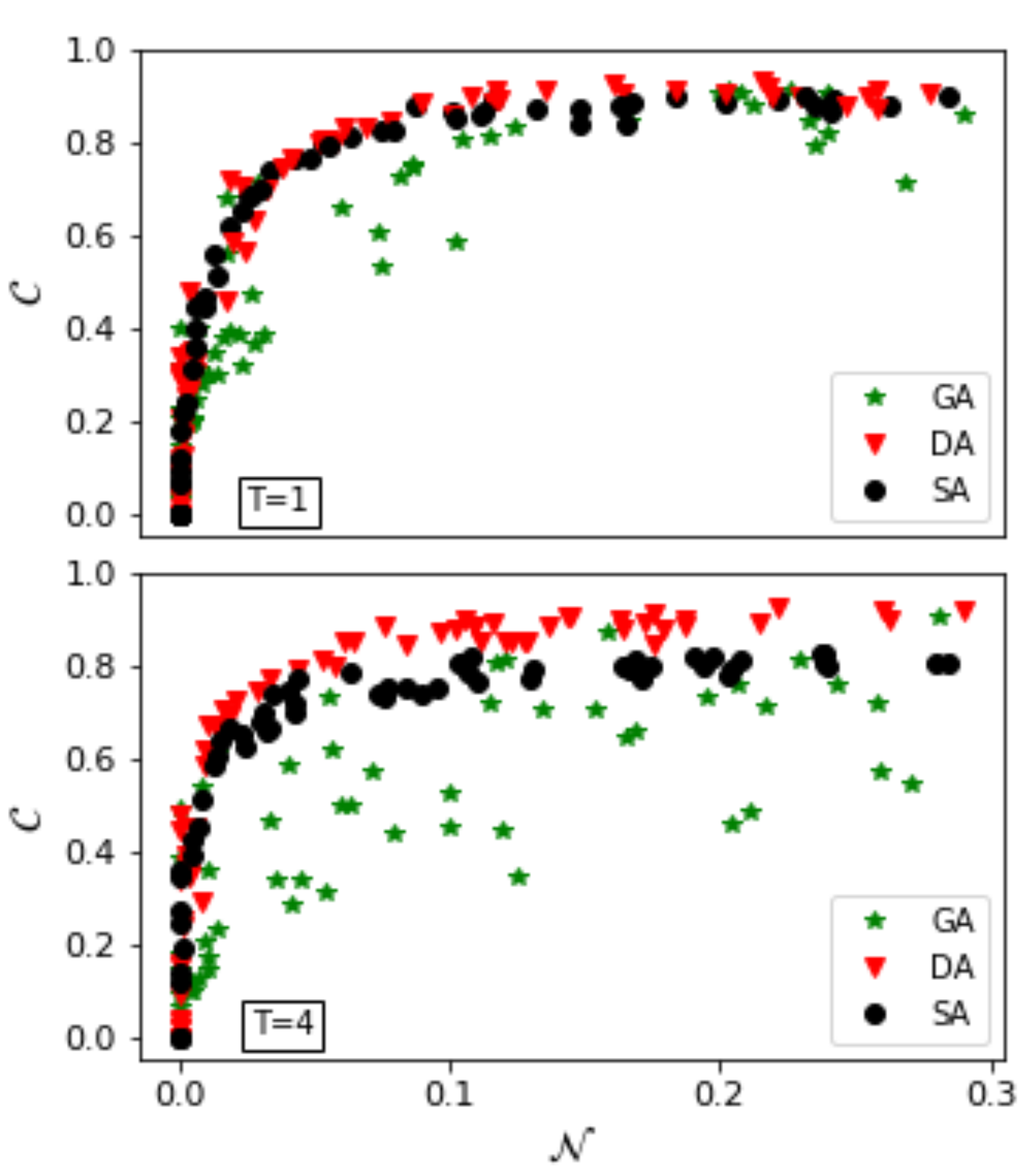}
\begin{footnotesize}
\caption{Optimal entanglement obtained as a function of the original degree of NM for the case of \textit{Single Addressing} (SA), \textit{Double Addressing} (DA) and \textit{Global Addressing} (GA). In the upper panel driving time is fixed as T=1, while in the lower panel T=4. Each point represents a physical situation in which random coupling constants $\alpha_1$ and $\alpha_2$ are chosen within the intervals $0<\alpha_{1}<1/\sqrt{2}$ and $0<\alpha_{2}<0.2$. Just points inside the interval $10^{-6}<\mathcal{N}<0.3$ are shown, while points with $\mathcal{N}<10^{-6}$ were excluded in the plot for being considered inside the Markovian regime. For further discussion about the Markovian regime, see main text and Fig. \ref{nm_con_y_sin_con}.}
\label{fig_res}
\end{footnotesize}
\end{center}
\end{figure}

For performing a quantitative analysis, we need to resort to numerical optimization for Eqs. (\ref{eqc1}) and (\ref{eqc2}) in order to maximize the functional given in Eq. (\ref{eq_conc}). Standard optimization tools from the Python SciPy library were used \cite{bib:scipy}. The procedure followed was to optimize over 10 different initial random seeds and to divide the driving time T into equidistant intervals such that the optimal fields possess 16 different amplitudes. It was seen that the optimal values for the concurrence did not improve adding more field amplitudes. In regards to the coupling constants of atom 1 and 2 with the environment, these were chosen randomly within some proper intervals ($0<\alpha_{1}<1/\sqrt{2}$ and $0<\alpha_{2}<0.2$) in order to ensure a proper resolution of the dynamics. In Fig. \ref{fig_res} we show the results obtained for the three different protocols analyzed (\textit{Single}, \textit{Double} and \textit{Global} \textit{Addressing}), by plotting the optimal concurrence obtained in each optimization as a function of the original degree of NM, quantified by the BLP measure. Points with $\mathcal{N}<10^{-6}$ were considered as Markovian points and were omitted from the plot. We will analyze these points separately.

\renewcommand{\figurename}{Figure} 

Interestingly, depending on which entangling protocol is being analyzed, a completely different behaviour of the entanglement as a function of NM arises. In first place, as was qualitatively suggested before, under the frame of \textit{Global Addressing} control there is no relation between a fix value of NM and the optimal concurrence obtained, i.e. the same amount of NM does not lead to the same amount of optimal concurrence. But surprisingly, both in the cases of \textit{Single} and \textit{Double Addressing} control the entangling fidelity is a direct function of the original degree of NM, independently of the coupling constant for each atom. This means that given a fixed value for the degree of NM of the free dynamics, there is a specific value for the maximum entanglement you could get by controlling the atoms. 

In the same way, as can be seen from Fig. \ref{fig_res}, there is a commitment between the benevolent and detrimental effects of the environment. While on one hand NM results the fundamental resource for generating entanglement, on the other hand, as time grows, decoherence worsens the optimal degree of entanglement that can be achieved by the control. This detrimental effect can be better appreciated for the case of \textit{Single Addressing} control, while the \textit{Double Addressing} method seems more robust to it. This is consistent with the fact that in the latter we have a greater degree of control over the system, since the both drivings are acting simultaneously.

\begin{figure}[!htb]
\begin{center}
\includegraphics[scale=0.63]{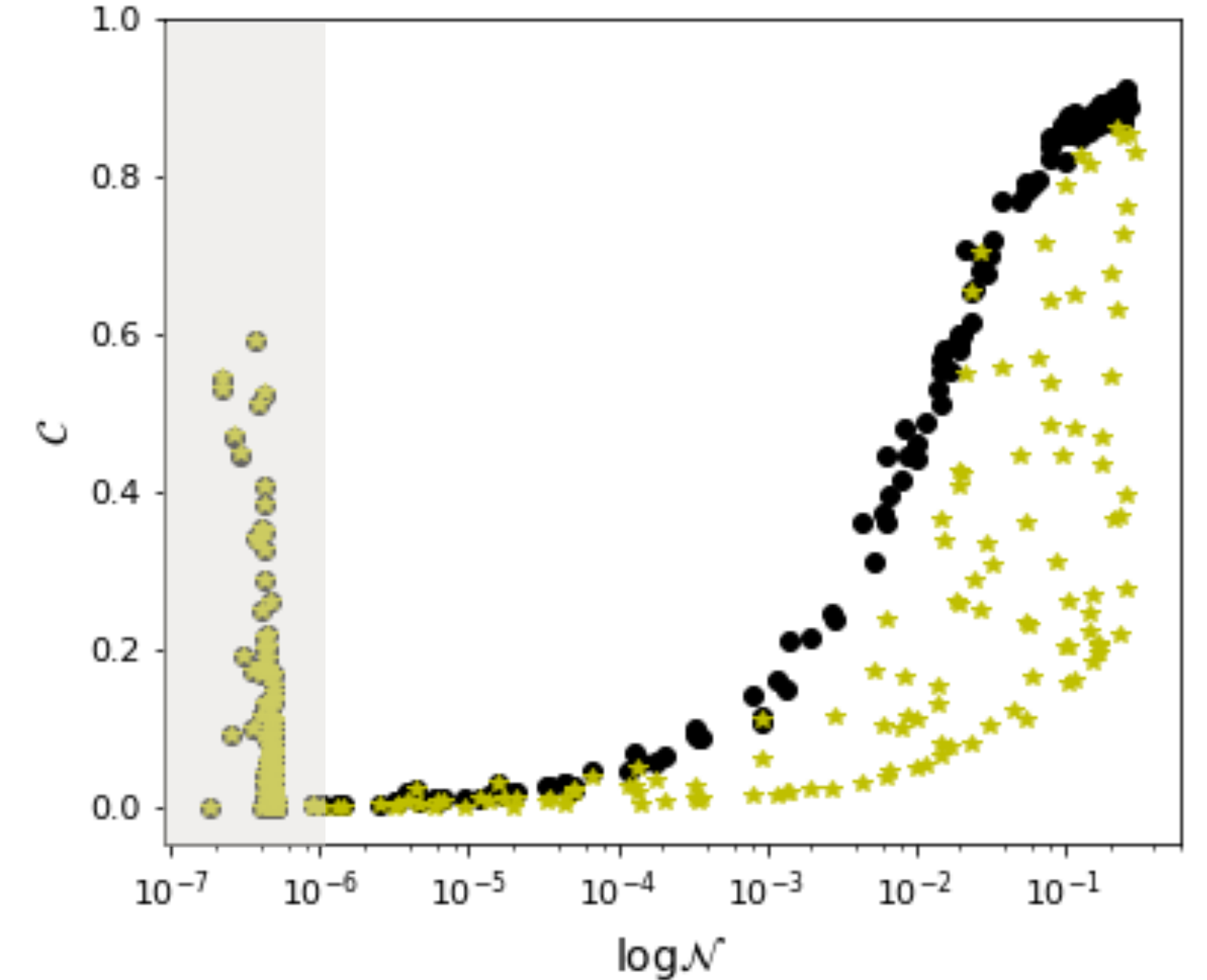}
\begin{footnotesize}
\caption{Concurrence as a function of the original degree of NM in the case without control (yellow stars) and considering control under the \textit{Single Addressing} method (black dots). Both Markovian and non-Markovian regions are shown for the same set of random couplings used in Fig. \ref{fig_res}. The Markovian regime is highlighted with a grey area. Driving time is T=1.}
\label{nm_con_y_sin_con}
\end{footnotesize}
\end{center}
\end{figure}

In order to shed more light into the role of NM under this entangling scheme, let us address now the following unanswered question: what is the role of the control in the Markovian regime? To answer this issue, let us take into consideration just the case of \textit{Single Addressing} control and analyze both the Markovian and non-Markovian regions. In Fig. \ref{nm_con_y_sin_con} we plot the concurrence as a function of NM and show not only the optimal degree of entanglement reached by the control, but also the natural degree of entanglement achieved when the dynamics is not being controlled externally.

\renewcommand{\figurename}{Figure} 

Notably, it can be seen that in the Markovian regime ($\mathcal{N}<10^{-6}$), there are some points in which despite having an appreciable amount of entanglement in the non-controlled case, the control field cannot do anything to improve their values considering there is no flow of information that allows to do so. The driving time is low enough in these cases so that revivals of distinguishability do not appear in the derivative of the trace distance (see Eq. (\ref{distinguibilidad})). On the other hand, the fact of having a certain amount of entanglement in this region is not surprising since it is well known that a common environment can entangle two subsystems coupled to it, independently of the existence of NM \cite{bib:ent_markov1,bib:ent_markov2}. However, the surprising fact is that in the non-Markovian region, given a specific amount of NM, this specific amount enables the control to generate a specific amount of entanglement. Indeed, in this regime we can observe that while the same amount of NM leads to different values of entanglement in the non-controlled case (yellow stars), the use of the control field generates a non-trivial increasing curve between entanglement and NM (black dots), strongly suggesting that the optimal amount of entanglement you finally get is a direct function of the original amount of NM in the system dynamics.

\subsection{Finite small-size environment: Spin star configuration}
\begin{figure}[!htb]
\begin{center}
\includegraphics[scale=0.60]{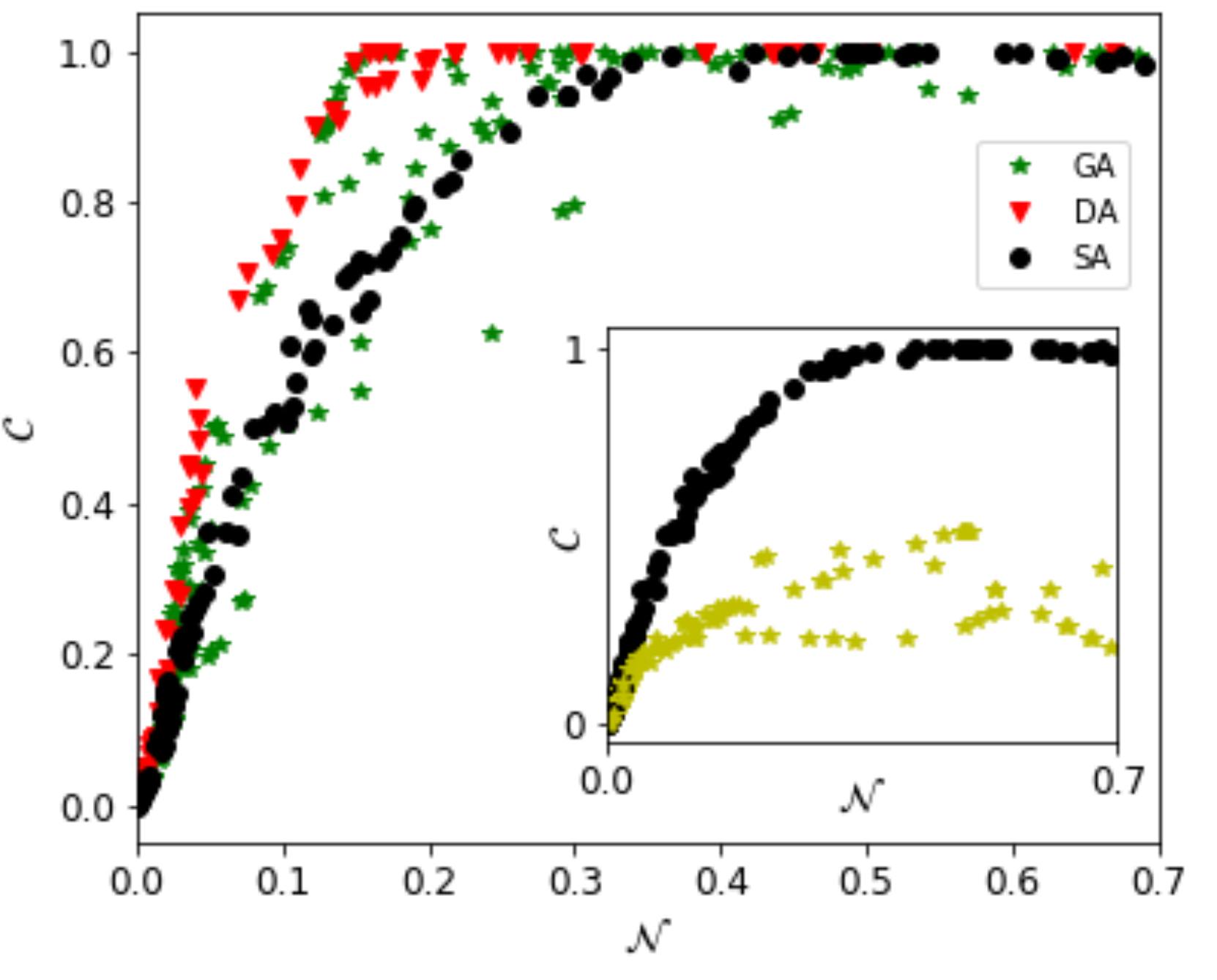}
\begin{footnotesize}
\caption{Concurrence as a function of the original degree of NM in the case of \textit{Single Addressing} (SA), \textit{Double Addressing} (DA) and \textit{Global Addressing} (GA). Each point represents a dynamics in which the coupling constants are chosen randomly between the intervals $0\leq A \leq 0.2$, the total number of spins are also arbitrarily chosen within $2\leq N \leq 8$ and the driving time is fixed in $T=10$. On the other hand, in the inset we show the concurrence as a function of the original degree of NM in the case without control (yellow stars) and considering control under the \textit{Single Addressing} method (black dots). The parameters in the inset are the same as in the main plot.}
\label{fig_spin_star}
\end{footnotesize}
\end{center}
\end{figure}

We now focus on a totally different system, where the size of the environment is finite and quite small. Indeed, we restrict ourselves to the case of $2\leq N \leq 8$, where we have complete knowledge over the whole open system plus environment. Thus, even though we are just interested in the open dynamics of the two non-interacting central spins, we solve the whole unitary optimized evolution and then trace over the environmental degrees of freedom to obtain the sought reduced dynamics. In Fig. \ref{fig_spin_star} we show the results for the concurrence as a function of the original amount of NM of the system dynamics for the three entangling protocols under consideration, i.e. \textit{Single}, \textit{Double} and \textit{Global Addressing}. The driving time has been divided into 250 equidistant time intervals, enough to ensure a proper resolution of the dynamics and the optimal concurrence shown is the maximum obtained after optimizing over 10 different initial random seeds. For this system, optimization tools from the open-source Python library QuTiP were used \cite{bib:qutip}.

\renewcommand{\figurename}{Figure} 

As in the system previously analyzed, we can also observe that both in the cases of \textit{Single} and \textit{Double Addressing} control the optimal entanglement reachable by the optimization is a direct function of the original amount of NM of the system dynamics, while in the case of \textit{Global Addressing} there is no quantitative relation at all. In the same way, it can be seen that given a fixed amount of NM, the \textit{Double Addressing} scheme allows to reach a better degree of entanglement, at least for low values of NM. On the other hand, in the inset of Fig. \ref{fig_spin_star} we show the natural degree of entanglement reached without performing any optimization as well as the optimal entanglement obtained in the case of \textit{Single Addressing} control. From here we can note that despite the entanglement in the non-controlled case is not determined by the degree of NM, the clever use of the control field generates again a non-trivial increasing curve between the optimal degree of entanglement and the degree of NM. As well as in the system of two non-interacting atoms in a leaky cavity, this is shown just for \textit{Single Addressing} control, but the same situation occurs in the \textit{Double Addressing} scheme. 

The novel analysis performed in this subsection for the spin star configuration allows us to extend and provide a deeper insight to the results obtained in Ref. \cite{bib:mirkin_ent} for a similar configuration, where other target states and cases were covered under \textit{Single Addressing} control.   

\section{Concluding remarks}
In this work we have sought for a quantitative and universal relation between entanglement and NM, if any. For this purpose, by using quantum optimal control as a searching tool, we have related the degree of success of a particular set of entangling tasks with respect to the degree of NM of the system dynamics. In this sense, by considering a physical setup composed of two non-interacting subsystems coupled to the same non-Markovian environment, we have revealed the existence of an entangling control task that is unachievable in the Markovian regime and whose degree of success depends univocally on the degree of NM of the dynamics. Such as to test the universality of this resource mechanism, two radically different systems were analyzed, on one hand a system composed of two non-interacting atoms coupled to the same infinite structured reservoir of harmonic oscillators and on the other hand a configuration of two non-interacting central spins coupled to the same small set of environmental spins. Despite the different nature of the systems considered, both cases rendered the same results, i.e. the degree of optimal entanglement reached by the optimization proved to be a direct function of the degree of NM of the system dynamics. In summary, the work consists in a practical demonstration of NM being exploited as a quantifiable and essential resource for generating entanglement in a general physical setup.

\begin{acknowledgements}
The work was partially supported by CONICET (PIP 112201 50100493CO), UBACyT (20020130100406BA), ANPCyT (PICT-2016-1056), and National Science Foundation (Grant No. PHY-1630114).
\end{acknowledgements}
\appendix
\section{Derivation of the reduced controlled dynamics of two non-interacting atoms in a leaky cavity}
\label{appA}
In order to derive the full reduced controlled dynamics given by the Hamiltonian of Eq. (\ref{eq_hamilt}), we need to go to the interaction representation with respect to $H_{S}$ and $H_{E}$ by means of an unitary transformation $U_{0}$ and obtain the transformed interaction Hamiltonian $\tilde{H}_{int}=U_{0}^{\dagger}H_{int}U_{0}$, where $U_{0}(t)=\Big( U_{0}^{(S_{1})}\otimes U_{0}^{(S_{2})} \Big)\otimes U_{0}^{(B)}(t)$. Assuming $\hbar=1$ from now on, we have
\begin{equation}
U_{0}^{(S_{j})}=exp \left(-i\int_{0}^{t}\omega_{j}(t')\sigma_{+}^{(j)}\sigma_{-}^{(j)}dt' \right)
\end{equation}
and 
\begin{equation}
    U_{0}^{(B)}(t)=exp \left(-i\sum_{k}\omega_{k}b^{\dagger}_{k}b_{k}t \right),
\end{equation}
which gives 
\begin{equation}
\begin{split}
    \tilde{H}_{int}(t) & =\left( \alpha_{1}\tilde{\sigma}_{+}^{(1)}(t)+\alpha_{2}\tilde{\sigma}_{+}^{(2)}(t) \right ) \otimes \tilde{B}(t).
\end{split}
\label{inth}
\end{equation}
We have defined $\tilde{B}(t)$ as  
\begin{equation}
    \tilde{B}(t)=U_{0}^{(B)\dagger}(t)\left(\sum_{k}g_{k}b_{k}\right)U_{0}^{(B)}(t)=\sum_{k}g_{k}e^{-i\omega_{k}t}b_{k}
\end{equation}
and 
\begin{equation}
\tilde{\sigma}_{+}^{(j)}(t)=U_{0}^{(S_{j})\dagger}(t)\sigma_{+}^{(j)}U_{0}^{(S_{j})}(t)=\sigma_{+}^{(j)}e^{iv_{j}(t)},
\end{equation}
where $v_{j}(t)=\int_{0}^{t}\omega_{j}(t')dt'$.

Starting with an initial state of the form
\begin{equation}
\ket{\phi(0)}=\left(C_{01}\ket{10}+C_{02}\ket{01} \right)\otimes_{k}\ket{0_{k}},
\end{equation}
the time evolution of the total system is given by
\begin{equation}
\begin{split}
\ket{\phi(t)} &=\ C_1(t)\ket{10}\ket{0_{B}}+C_2(t)\ket{01}\ket{0_{B}} \\ & +\sum_{k}C_{k}(t)\ket{00}\ket{1_{k}},
\end{split}
\end{equation} 
being $\ket{1_{k}}$ the state of the reservoir with only one excitation in the k-th mode. If we note $\ket{\varphi_{10}}=\ket{10}\ket{0_{B}}$, $\ket{\varphi_{01}}=\ket{01}\ket{0_{B}}$, $\ket{\varphi_{k}}=\ket{00}\ket{1_{k}}$ we then have 
\begin{equation}
    \ket{\dot{\phi}(t)}=\dot{C}_{1}\ket{\varphi_{10}}+\dot{C}_{2}\ket{\varphi_{01}}+\sum_{k}\dot{C}_{k}\ket{\varphi_{k}}.
\end{equation}
As we are interested in the equations of motion for the evolution of the coefficients $C_{j}(t)$, we must calculate the following elements
\begin{equation}
\begin{split}
&\ \bra{\varphi_{10}}\ket{\dot{\phi}}= \dot{C}_{1}=-i\bra{\varphi_{10}}\tilde{H}_{int}\ket{\phi(t)} \\ &
\braket{\varphi_{01}}{\dot{\phi}}=\dot{C}_{2}=-i\bra{\varphi_{01}}\tilde{H}_{int}\ket{\phi(t)} \\ &
\braket{\varphi_{k}}{\dot{\phi}}=\dot{C}_{k}=-i\bra{\varphi_{k}}\tilde{H}_{int}\ket{\phi(t)}.         
\end{split}
\end{equation}
In first place, 
\begin{equation}
\begin{split}
        \bra{\varphi_{10}}\tilde{H}_{int}\ket{\varphi_{10}}& = \ \bra{10}\alpha_{1}\tilde{\sigma}_{+}^{(1)}+\alpha_{2}\tilde{\sigma}_{+}^{(2)}\ket{10}\bra{0_{B}}\tilde{B}\ket{0_{B}}  \\ & +(...)\bra{0_{B}}\tilde{B}^{\dagger}\ket{0_{B}},
\end{split}
\end{equation}
but $\bra{0_{B}}\tilde{B}\ket{0_{B}}$ and $\bra{0_{B}}\tilde{B}^{\dagger}\ket{0_{B}}$ are both zero. The same happens for elements $\bra{\varphi_{01}}\tilde{H}_{int}\ket{\varphi_{01}}$ and for $\bra{\varphi_{10}}\tilde{H}_{int}\ket{\varphi_{01}}$. On the other hand, 
\begin{equation}
\begin{split}
\bra{\varphi_{10}}\tilde{H}_{int}\ket{\varphi_{k}} & = \ \bra{10}\alpha_{1}\tilde{\sigma}_{+}^{(1)}+\alpha_{2}\tilde{\sigma}_{+}^{(2)}\ket{00}\bra{0_{B}}\tilde{B}\ket{1_{k}} \\ & + \bra{10}\alpha_{1}^{*}\tilde{\sigma}_{-}^{(1)}+\alpha_{2}^{*}\tilde{\sigma}_{-}^{(2)}\ket{00}\bra{0_{B}}\tilde{B}^{\dagger}\ket{1_{k}} \\ & = \alpha_{1}e^{iv_{1}(t)}\bra{10}\ket{10}g_{k}e^{-i\omega_{k}t}. 
\end{split}
\end{equation}
In consequence 
\begin{equation}
\dot{C}_{1}=-i\alpha_{1}\sum_{k}g_{k}e^{i\left(v_{1}(t)-\omega_{k}t \right)}C_{k}(t),
\label{eq_c1}
\end{equation}
and for symmetry
\begin{equation}
\dot{C}_{2}=-i\alpha_{2}\sum_{k}g_{k}e^{i\left(v_{2}(t)-\omega_{k}t \right)}C_{k}(t).
\end{equation}
In an analogous way, 
\begin{equation}
\begin{split}
&\ \bra{\varphi_{k}}\tilde{H}_{int}\ket{\varphi_{k}}\propto \bra{0}\sigma_{\pm} \ket{0}=0 \\ &
\bra{\varphi_{k}}\tilde{H}_{int}\ket{\varphi_{10}}=g_{k}^{*}\alpha_{1}e^{-i(v_{1}(t)-\omega_{k}t)} \\ &
\bra{\varphi_{k}}\tilde{H}_{int}\ket{\varphi_{01}}=g_{k}^{*}\alpha_{2}e^{-i(v_{2}(t)-\omega_{k}t)}.
\end{split}
\end{equation}
Consequently, 
\begin{equation}
\begin{split}
\dot{C}_{k}(t) & = \ -ig_{k}^{*} \Big( \alpha_{1}e^{-i(v_{1}(t)-\omega_{k}t)}C_{1}(t) \\ & +\alpha_{2}e^{-i(v_{2}(t)-\omega_{k}t)}C_{2}(t) \Big).
\end{split}
\end{equation}
Integrating the last equation, 
\beq
\begin{split}
C_{k}(t)-C_{k}(0) & = \ -ig_{k}^{*}\int_{0}^{t}dt'\Big( \alpha_{1}e^{-i(v_{1}(t')-\omega_{k}t')}C_{1}(t')  \\ & +\alpha_{2}e^{-i(v_{2}(t')-\omega_{k}t')}C_{2}(t') \Big),
\end{split}
\label{eq_ck}
\eeq
where $C_{k}(0)=0$ since the initial bath state is vacuum. From now on we note $\Delta_{k}^{(j)}(t)=v_{j}(t)-\omega_{k}t$.

Inserting Eq. (\ref{eq_ck}) into Eq. (\ref{eq_c1}) we get 
\beq
\begin{split}
\dot{C}_{1}(t) &= \  -\sum_{k}|g_{k}|^{2}e^{i\Delta_{k}^{(1)}(t)}\int_{0}^{t}dt' \Big( \alpha_{1}^{2}e^{-i\Delta_{k}^{(1)}(t')}C_{1}(t') \\ & +\alpha_{1}\alpha_{2}e^{-i\Delta_{k}^{(2)}(t')}C_{2}(t') \Big) \\ & =-\int_{0}^{t}dt'\sum_{k}|g_{k}|^{2} \Big(\alpha_{1}^{2}e^{i(\Delta_{k}^{(1)}(t)-\Delta_{k}^{(1)}(t'))}C_{1}(t') \\ & +\alpha_{1}\alpha_{2}e^{i(\Delta_{k}^{(1)}(t)-\Delta_{k}^{(2)}(t'))}C_{2}(t') \Big),
\end{split}
\eeq
where $\Delta_{k}^{(a)}(t)-\Delta_{k}^{(b)}(t')=v_{a}(t)-\omega_{k}t-v_{b}(t')+\omega_{k}t'=-\omega_{k}(t-t')+v_{a}(t)-v_{b}(t')$. 

In the continuum limit for the environment, we can introduce a general spectral density $J(\omega)$ and obtain
\begin{equation}
\begin{split}
\dot{C}_{1}(t) &=\-\int_{0}^{t}dt' \left(\int d\omega J(\omega)e^{-i\omega(t-t')} \right) \\ &  \Big(\alpha_{1}^{2}e^{i(v_{1}(t)-v_{1}(t'))}C_{1}(t') \\ & +\alpha_{1}\alpha_{2}e^{i(v_{1}(t)-v_{2}(t'))}C_{2}(t') \Big).
\end{split}
\end{equation}
Considering that $F(t-t')=\int d\omega J(\omega)e^{-i\omega(t-t')}$ is the bath correlation function, the above equation then becomes
\begin{equation}
\begin{split}
\dot{C}_{1}(t) &=\-\int_{0}^{t}dt'e^{iv_{1}(t)}F(t-t') \\ &  \Big(\alpha_{1}^{2}e^{-iv_{1}(t')}C_{1}(t') +\alpha_{1}\alpha_{2}e^{-iv_{2}(t')}C_{2}(t') \Big).
\end{split}
\label{eq_c1v2}
\end{equation}
And for symmetry we have
\begin{equation}
\begin{split}
\dot{C}_{2}(t) &=\-\int_{0}^{t}dt'e^{iv_{2}(t)}F(t-t') \\ &  \Big(\alpha_{2}^{2}e^{-iv_{2}(t')}C_{2}(t') +\alpha_{1}\alpha_{2}e^{-iv_{1}(t')}C_{1}(t') \Big).
\end{split}
\label{eq_c2v2}
\end{equation}
At this point is interesting to note that the equations for $C_{1}(t)$ and for $C_{2}(t)$ are both coupled with each other, which suggests us the possibility that the control field may be able to tune the entanglement between both atoms. \\
Now let us rewrite Eqs. (\ref{eq_c1v2}) and (\ref{eq_c2v2}) as
\begin{equation}
    \dot{C}_{l}(t)=-\int_{0}^{t}dt'F(t-t')e^{iv_{l}(t')}m_{l}(t'),
\end{equation}
where $l=1,2$. Considering $J(\omega)$ to be of Lorentzian form, i.e. 
\begin{equation}
    J(\omega)=\dfrac{\gamma_{0}}{2\pi} \dfrac{\lambda^{2}}{(\omega-\omega_{0})^{2}+\lambda^{2}},
\end{equation}
where $\gamma_{0}$ refers to the effective coupling between the system and the bath and $\lambda$ is the width of the spectral density. In this frame, we get for $F(t-t')$ 
\begin{equation}
    F(t-t')=p_{0}e^{q_{0}(t-t')},
\end{equation}
where we have defined $p_{0}=\dfrac{\gamma_{0}\lambda}{2}$ and $q_{0}=-(\lambda+i\omega_{0})$. Consequently, by defining $g_{1}(t)=\dot{C}_{1}(t)/C_{1}(t)$, we have
\begin{equation}
    g_{1}(t)=\dfrac{\dot{C}_{1}(t)}{C_{1}(t)}=-\int_{0}^{t}dt'F(t,t')e^{iv_{1}(t)}\dfrac{m_{1}(t')}{C_{1}(t)}.
    \label{eq_g1}
\end{equation}
If we now call $h(t,t')=e^{iv_{1}(t)}\dfrac{m_{1}(t')}{C_{1}(t)}$, then 
\begin{equation}
    \dfrac{\partial}{\partial t}h(t,t')=m_{l}(t') \Big(\dfrac{i\dot{v}_{l}(t)C_{1}(t)-\dot{C}_{1}(t)}{C_{1}^{2}(t)} \Big) e^{iv_{l}(t)}
\end{equation}
\begin{equation}
\dfrac{\partial}{\partial t}h(t,t')=h(t,t')(i\omega_{1}(t)-g(t)),
\end{equation}
and so we have
\begin{equation}
\begin{split}
\dot{g}_{1}(t)&=\ -\dfrac{d}{dt} \Big(\int_{0}^{t}dt'F(t,t')h(t,t') \Big) \\&
= -F(0)h(t,t)-\int_{0}^{t}dt'\Big(\dfrac{\partial}{\partial t} (F(t,t')h(t,t')) \Big) \\ &
= -p_{0}h(t,t)-\int_{0}^{t}dt' \big(q_{0}F(t,t')h(t,t') \\ & +F(t,t')\dfrac{\partial}{\partial t}h(t,t') \big) \\ &  =-p_{0}h(t,t)+q_{0}g_{1}(t)+(i\omega_{1}(t)-g_{1}(t))g_{1}(t) \\ & =-p_{0}h(t,t)+g_{1}(t)(q_{0}+i\omega_{1}(t)-g_{1}(t)) \\ & =-p_{0}e^{iv_{1}(t)}\dfrac{m_{1}(t)}{C_{1}(t)}+g_{1}(t)(q_{0}+i\omega_{1}(t)-g_{1}(t)) \\ & =-p_{0}(\alpha_{1}^{2}+\alpha_{1}\alpha_{2}\dfrac{C_{2}(t)}{C_{1}(t)}e^{i(v_{1}(t)-v_{2}(t))}) \\ & +g_{1}(t)(q_{0}+i\omega_{1}(t)-g_{1}(t)) \\ & = -p_{0}\alpha_{1} \Big(\alpha_{1}+\alpha_{2}\dfrac{C_{2}(t)}{C_{1}(t)}e^{i(v_{1}(t)-v_{2}(t))}\Big) \\ & +g_{1}(t) (q_{0}+i\omega_{1}(t)-g_{1}(t)).
\end{split}
\end{equation}
Considering that at the same time, by differentiating Eq. (\ref{eq_g1}), we get 
\begin{equation}
    \dot{g}_{1}=\dfrac{\Ddot{C}_{1}C_{1}-\dot{C}_{1}^{2}}{C_{1}^{2}},
\end{equation}
then 
\begin{equation}
\begin{split}
\dfrac{\Ddot{C}_{1}}{C_{1}}-\dfrac{\dot{C}_{1}^{2}}{C_{1}^{2}} & = \ -p_{0}\alpha_{1}\Big(\alpha_{1}+\alpha_{2}\dfrac{C_{2}(t)}{C_{1}(t)}e^{i(v_{1}-v_{2})} \Big) \\ & + \dfrac{\dot{C}_{1}}{C_{1}}\Big(q_{0}+i\omega_{1}-\dfrac{\dot{C}_{1}}{C_{1}}\Big).
\end{split}
\label{eq_difc1}
\end{equation}
Noting that $p_{0}=\gamma_{0}\lambda/2$, $\omega_{1}(t)=\omega_{0}+\epsilon_{1}(t)$ and $q_{0}=-\lambda-i\omega_{0}$, we obtain $q_{0}+i\omega_{1}=-\lambda-i\omega_{0}+i\omega_{0}+i\epsilon_{1}(t)=i\epsilon_{1}(t)-\lambda$. The differential Eq. (\ref{eq_difc1}) then becomes
\begin{equation}
    \Ddot{C}_{1}=-(\lambda-i\epsilon_{1}(t))\dot{C}_{1}-\alpha_{1}\dfrac{\gamma_{0}\lambda}{2}\Big(\alpha_{1}C_{1}+\alpha_{2}e^{i(v_{1}-v_{2})}C_{2}\Big).
\end{equation}
Reordering the terms, we finally get 
\begin{equation}
    \Ddot{C}_{1}+(\lambda-i\epsilon_{1}(t))\dot{C}_{1}+\alpha_{1}\dfrac{\gamma_{0}\lambda}{2}\Big(\alpha_{1}C_{1}+\alpha_{2}e^{i(v_{1}-v_{2})}C_{2}\Big)=0.
\end{equation}
And for symmetry
\begin{equation}
        \Ddot{C}_{2}+(\lambda-i\epsilon_{2}(t))\dot{C}_{2}+\alpha_{2}\dfrac{\gamma_{0}\lambda}{2}\Big(\alpha_{2}C_{2}+\alpha_{1}e^{-i(v_{1}-v_{2})}C_{1}\Big)=0,
\end{equation}
where $v_{l}(t)=\int_{0}^{t}ds \omega_{l}(s)$. These are the exact equations of motion for the reduced dynamics of the driven two non-interacting atoms coupled to a Lorentzian bath at zero temperature, under the Rotating Wave Approximation (RWA).

\clearpage
\bibliography{main.bib}

\end{document}